\documentclass[11pt]{cernrep}

\usepackage{graphicx}
\usepackage{bm} 
\usepackage{here}

\def\slashchar#1{\setbox0=\hbox{$#1$}
   \dimen0=\wd0 \setbox1=\hbox{/} \dimen1=\wd1
   \ifdim\dimen0>\dimen1 \rlap{\hbox to \dimen0{\hfil/\hfil}} #1
   \else  \rlap{\hbox to \dimen1{\hfil$#1$\hfil}} / \fi}

\begin{document}

\title{\bf IMPACT-PARAMETER DEPENDENCE OF THE DIAGONAL GPD
OF THE PION FROM CHIRAL QUARK MODELS\thanks{
Presented at ``Light Cone Workshop: HADRONS AND BEYOND'', 5-9 August 2003, University of Durham (U.K.)}} 
\author{\underline{Wojciech Broniowski}$^1$ and Enrique Ruiz Arriola$^2$} 
\institute{$^1$The H. Niewodnicza\'nski Institute of Nuclear Physics,
Polish Academy of Sciences, \\~~~PL-31342 Cracow, Poland\\
$^2$Departamento de F\'{\i}sica Moderna, Universidad de Granada, E-18071 Granada, Spain}

\maketitle

\begin{abstract}
We analyze the impact-parameter dependent diagonal non-singlet generalized 
parton distribution of the pion in two distinct chiral 
quark models: the Spectral Quark Model and the
Nambu-Jona-Lasinio model. Leading-order perturbative QCD evolution from the low quark-model scale 
to higher scales is carried out with the help of the inverse Mellin transform. The
model agrees very reasonably with the recent results from transverse-lattice calculations,
while the forward parton distribution agrees with the experimental data.
\end{abstract}

This talk is based on our recent work \cite{us}, where more details 
and references can be found.
Recently, transverse-lattice calculations have produced first data
\cite{Dalley:2003sz} on the {\em impact-parameter dependent} diagonal
non-singlet generalized parton distributions (GPD) of the pion ({\em cf.} contribution by Dalley).
These GPD's have been a subject of intense studies 
(for a review and literature see, {\em e.g.}, Ref.~\cite{max,diehl}).
The impact-parameter ($b$) formulation, discussed here by Burkardt, 
is a basis for the transverse lattices 
\cite{Dalley:2002nj}, as shown by van de Sande.
In this talk we present predictions for the GPD of the pion from
two chiral quark models: the recently-proposed {\em Spectral
Quark Model} (SQM)~\cite{RuizArriola:2003bs} and the {\em
Nambu--Jona-Lasinio} model with the Pauli-Villars
regulator (NJL). 
Chiral model results for the skewed $b$-integrated (forward) GPD \cite{TNV} were shown here by Vento. 

\begin{figure}[tb]
\begin{center}
\includegraphics[width=14cm]{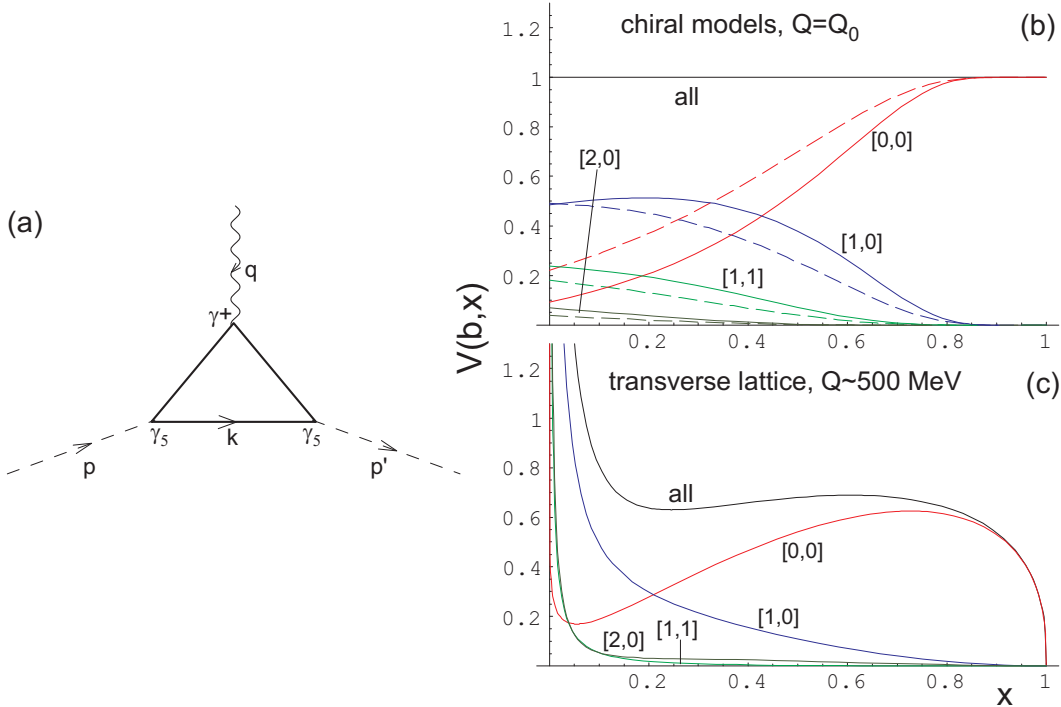}
\end{center} 
\caption{(a) The diagram for the evaluation of the GPD of the pion in chiral quark
models. (b) Valence impact-parameter dependent diagonal 
GPD of the pion, $V(x,b)$, plotted as a function of the Bjorken $x$ variable.
The results of the chiral quark models at the model scale of $Q=Q_0=313$~MeV. Solid lines: 
SQM of Ref.~\cite{RuizArriola:2003bs}, dashed lines: the
NJL model with two Pauli-Villars subtractions. Label
{\em all} denotes the forward distribution, {\em i.e.}, the function
$V(x,b)$ integrated over the whole $b$-plane. Labels $[i,j]$ denote
the function $V(x,b)$ integrated over the square plaquettes centered
at coordinates $(i b_0,j b_0)$ of the edge of length $b_0$, times the
degeneracy of the plaquette (see the text for details). Following
Ref.~\cite{Dalley:2003sz}, the value of $b_0$ is taken to be $2/3$~fm.
(c) The results for $V(x,b)$ at the scale $Q \sim 500$~MeV, obtained
from transverse-lattice calculation of
Ref.~\cite{Dalley:2003sz}. The model results of (b) can be compared to the data of
(c) only after a suitable QCD evolution.}
\label{fig:fig2}
\end{figure}

The off-forward ($\bf{\Delta_ \perp} \neq 0$) diagonal ($\xi=0$)
GPD of the pion (we take $\pi^+$) is defined as
\begin{eqnarray}
H(x,\xi = 0, - {\bf{\Delta}}_ \perp^2 ) = \int d^2 b \int
\frac{dz^-}{4\pi} e^{i ( x p^+ z^- + {\bf{\Delta}}_\perp \cdot
{\bf{b}} )} 
\langle \pi^+ (p') | \bar q (0,
-\frac{z^-}{2} , {\bf{b}} ) \gamma^+ q (0, \frac{z^-}{2} , {\bf{b}} )
| \pi^+ (p) \rangle ,
\end{eqnarray} 
where $x$ is the Bjorken $x$, and $\Delta_\perp=p'-p$ lies in the
transverse plane.  
In chiral quark models the evaluation of $H$ at the leading-$N_c$
(one-loop) level amounts to the calculation of the diagram of
Fig.~\ref{fig:fig2}(a),
where the solid line denotes the propagator of the quark of mass
$\omega$. The calculation is carried in the light-cone coordinates and in the Breit frame, $\Delta^+
=0$~\cite{Frederico:ye}, yielding in the chiral limit 
\begin{eqnarray}
H(x,0,-{\bf \Delta}_\perp^2; \omega) = \frac{N_c \omega^2}{\pi
f_\pi^2} \int \frac{d^2 {\bf K}_\perp}{(2\pi)^2} \frac{ 1 +
\frac{ {\bf K}_\perp \cdot \Delta_\perp (1-x)}{{\bf K}_\perp^2 +
\omega^2}}{({\bf K}_\perp+(1-x)\Delta_\perp)^2 + \omega^2},
\label{eq:unint_ff}
\end{eqnarray} 
where $ {\bf K}_\perp=(1-x){\bf p}_\perp - x {\bf k}_\perp$.
To proceed further, we need to specify the regularization. First, we
consider the recently proposed {\em Spectral Quark Model} (SQM)
\cite{RuizArriola:2003bs}, described in ERA's talk. 
The approach is very successful in
describing both the low- and high-energy phenomenology of the pion, while complying
to the chiral symmetry and anomalies. In the meson dominance version of the model
one obtains \cite{us}
\begin{eqnarray} 
H(x,0,- {\bf \Delta}_\perp^2) = \frac{m_\rho^2 ( m_\rho^2 - (1-x)^2
{\bf \Delta}_\perp^2)} {( m_\rho^2 + (1-x)^2 {\bf
\Delta}_\perp^2)^2}. \label{res:vmd}
\end{eqnarray} 
We check that $H(x,0,0)=1 $ and $\int_0^1 dx
H(x,0,t) = m_\rho^2/(m_\rho^2+t)$, which is the
built-in vector-meson dominance principle. 
The NJL model with the Pauli-Villars regularization 
also yields simple expressions.

Our aim is to compare our results, after a suitable QCD evolution,
to the transverse-lattice data of Ref.~\cite{Dalley:2003sz}. These
data give the GPD of the pion at
discrete values of the impact parameter ${\bf b}$, corresponding to a
square lattice with spacing of $b_0\simeq 2/3$~fm. 
In order 
to mimic the lattice in our study,
the model predictions are {\em smeared over
square plaquettes}, the same ones as in the discrete lattice. The plaquettes are labeled $[i,j]$, 
which means that they are centered at coordinates
$(i b_0,j b_0)$, and have the edge of length $b_0=2/3$~fm \cite{Dalley:2003sz}. The smeared 
valence GPD is defined as
\begin{eqnarray}
V(x,[i,j])\equiv \int_{(i-1/2) b_0}^{(i+1/2) b_0} db_1 \int_{(j-1/2) b_0}^{(j+1/2) b_0} db_2 
V(x,\sqrt{b_1^2+b_2^2}).
\label{smear} 
\end{eqnarray}
Figure \ref{fig:fig2}(b) shows the model results with smearing. The degeneracy factor equal to
the number of plaquettes 
equidistant from the origin is included, {\em i.e.}, the $[1,0]$, $[1,1]$, and $[2,0]$ plaquettes are multiplied 
by a factor of four, while $[2,1]$ would be multiplied by eight. 

Figure 2(c) shows the data from the transverse-lattice calculations
shown by Dalley~\cite{Dalley:2003sz}.  They refer
to the scale $Q \simeq 500$~MeV.
Since the scale pertaining to our calculation is much lower, we
need to evolve our results upward before comparing to the data of Fig.~2(c).
Our model calculation has produced distributions
corresponding to a low quark model scale, $Q_0$. A way to estimate this scale is to run the QCD
evolution upward from various scales $Q_0$ up to a scale $Q$ where the
data can be used.  Alternatively, one may use the
momentum fraction carried by the quarks at the scale $Q$ and the
downward QCD evolution in order to estimate $Q_0$.
The resulting value is $Q_0=313^{+20}_{-10} {\rm MeV}$.

We apply the QCD evolution to the smeared
functions of Eq.~(\ref{smear}). 
Figure 2 (a) shows the plaquette-averaged functions $V(x,Q_0,[i,j])$
for SQM (solid lines) and the NJL model (dashed
lines). 
The results of the evolution for SQM (for NJL the effect is similar)
are shown in Fig.~3 at three
values of the reference scale $Q$: 400~MeV (a), 500~MeV (b), and 2~GeV
(c).  We note a large effect of the evolution on the distribution
functions. The lines labeled {\em all} correspond to the forward case,
{\em i.e.}, show 
\mbox{$\int d^2b \, V(x,Q,b)=V(x,Q,{\bf \Delta}_\perp=0)$}. The
originally flat distribution of Fig.~2(a) recovers the correct
end-point behavior at $x \to 1$ according to Eq.~(\ref{endpoint}). As
$Q$ increases, the distribution is pushed towards lower values of $x$,
as is well known for the DGLAP evolution. At $Q=2$~GeV the result
agrees very well~\cite{DR95} with the SMRS parameterization of the pion structure
function \cite{SMRS92}, as can be seen from Fig.~3(d) 
(here we plot  for convenience $x V(x,Q)$) by comparing the
dashed and solid lines.
\begin{figure}[tb]
\begin{center}
\includegraphics[width=14cm]{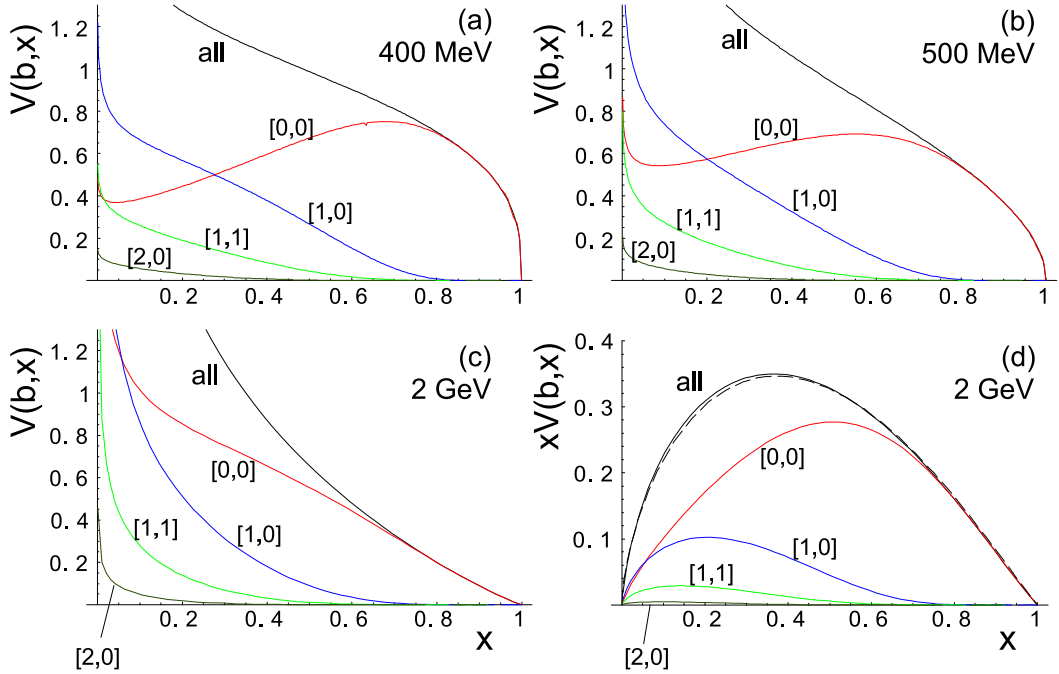}
\end{center} 
\caption{Results of the LO DGLAP evolution of the impact-parameter
dependent diagonal non-singlet GPD of the
pion, $V(x,b,[i,j])$, started from the initial condition at
$Q=Q_0=313$~MeV produced by the Spectral Quark Model (Fig.~2(a), solid
lines). Figures (a,b,c) correspond to $Q=400$~MeV, $500$~MeV, and
$2$~GeV, respectively. Labels as in Fig.~2. Figure (d) shows $x
V(x,b,[i,j])$ for $Q=2$~GeV, with the dashed line showing the SMRS
\cite{SMRS92} parameterization of the data for the forward parton distribution function.
}
\label{fig:evo}
\end{figure}
The qualitative agreement of Figs. 1(c) and 2(b) is striking, baring in mind the simplicity 
of our approach.

An interesting feature of LO DGLAP evolution is the induced
suppression at \mbox{$x \to 1$}: a function which
originally behaves as \mbox{$V(x,Q_0,b) \to C(b) (1-x)^p$} evolves
into \cite{Peterman}
\begin{eqnarray} 
V (x,Q,b) \to C(b) (1-x)^{p - \frac{4 C_F }{\beta_0} \log
\frac{\alpha(Q)}{ \alpha(Q_0) }}, \qquad x\to 1.
\label{endpoint}
\end{eqnarray} 
In our approach the integrated function at $Q=Q_0$ has $p=0$ and the exponent becomes 
$1.1 \pm 0.1$ at $Q=2$~GeV and $1.3 \pm 0.1$ at $Q=4$~GeV. Note that 
the Brodsky-Lepage counting rules for the behavior at $x \to 1$ are disobeyed, 
as well as the predictions are different from the Dyson-Schwinger model
presented here by Roberts, where the behavior is $(1-x)^2$. 
On the other hand, our predictions agree within
experimental uncertainties with 
the experimental data. Fig. 3 confronts our results evolved to the scale of 4~GeV to the
E615 experimental Drell-Yan data \cite{e615} which cover the large-$x$ region. 
The quality of this comparison is impressive.
\begin{figure}[tb]
\begin{center}
\includegraphics[width=7.7cm]{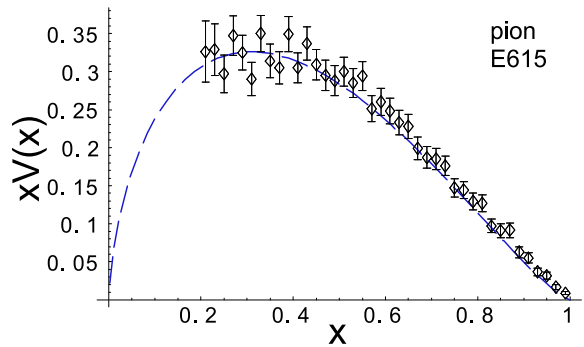}
\end{center} 
\caption{Model prediction for the valence forward parton distribution
of the pion, evolved to the scale of 4~GeV, and the E615 Drell-Yan data \cite{e615}.
The behavior at $x \to 1$ is $(1-x)^{1.3 \pm 0.1}$.}
\label{fig:e615}
\end{figure}

In summary, the obtained agreement of our approach, based on non-perturbative 
chiral quark models in conjunction with perturbative LO
DGLAP evolution, with the data from the transverse lattices, is quite
remarkable and encouraging, baring in mind the simplicity of the
models and the apparently radically different handling of chiral
symmetry in both approaches. We also note that the forward parton distribution
functions are in agreement with the SMRS parameterization and 
experimental Drell-Yan data.

We are grateful to Simon Dalley, Matthias Burkardt, and 
Brett van de Sande for helpful discussions concerning the
transverse-lattice data.
Supported in part by the Spanish
Ministerio de Asuntos Exteriores and the Polish State Committee for
Scientific Research, grant number 07/2001-2002, 
by the Spanish DGI, grant no. BFM2002-03218, and by Junta de
Andaluc\'{\i}a, grant no. FQM-225.

\end{document}